\documentclass[%
 aip,
 jmp,%
 amsmath,amssymb,
 reprint,%
]{revtex4-1}
\usepackage{comment}
\usepackage{graphicx}
\usepackage{dcolumn}
\usepackage{bm}
\usepackage{amsmath}
\usepackage{amsthm} 

\usepackage{caption}

 \usepackage{mathptmx} 

\begin{document}


\title{Biodiversity, extinctions and evolution
of ecosystems with shared resources
}


\author{Vladimir Kozlov}
\email[]{vladimir.kozlov@liu.se}
\affiliation{Dept.of Mathematics, University of Linkoping, 58183, Linkoping, Sweden}

\author{Sergey  Vakulenko}
\affiliation{Institute for Mechanical Engineering Problems, Russian Academy of Sciences,  Saint Petersburg, Russia
and  \\ Saint Petersburg National Research University of Information Technologies, Mechanics and Optics.
Saint Petersburg, Russia
}

\author{Uno Wennergren}
\affiliation{Dept.of Ecology, University of Linkoping, 58183, Linkoping, Sweden
}


\date{\today}

\begin{abstract}
We investigate the formation of stable ecological networks where many species share the same resource.
We show that  such stable ecosystem naturally occurs as a result of extinctions. We obtain an analytical relation for the number of coexisting species and find a relation describing how many species that may go extinct as a result of a sharp environmental change.  We introduce a special parameter that is a combination of species traits and resource characteristics used in the model formulation. This parameter describes the pressure on system to converge, by extinctions.  When that stress parameter is large we obtain that the species traits concentrate at some values. This stress parameter is thereby a parameter that determines the level of final biodiversity of the system. 
Moreover, we show that dynamics of this limit system can be described by  simple differential equations.
\end{abstract}


\maketitle


\section{Introduction}\label{Intro}

Thermodynamics and Statistical Physics allow us to describe equilibrium states of systems consisting of many particles by a few variables. This approach is very effective for many physical and chemical applications, and particularly, when we are dealing with closed systems. It would be very tempting to find such reduced macroscopic descriptions for
large ecological and economic systems. 

Recently, in the paper \cite{Barab} a very reduced description is proposed for a class of ecological models. It is based on an analog of the mean field theory and aims to describe possible bifurcations. In the present paper, we suggest a variant of a reduced description for other large class of ecosystems, where many species compete for a few resources.
Our second goal is to consider 
one of the most intriguing puzzles in ecosystem theory, namely,   the biodiversity problem: why can such a large number of similar species share the same habitat and how to estimate this number.
Aquatic ecosystems is a typical example where many different phytoplankton species do
coexist.  The
principle of competitive exclusion \cite{Volterra, Hardin}  asserts that different species sharing  the same
resource cannot coexist and it predicts that an assembly of competing species will converge to a single species.
In general, competition models  
show that the number of species that can coexist in
equilibrium cannot be greater than the number of limiting
factors \cite{Til77, Zeeman}.
However,  hundreds of species
of phytoplankton coexist although only nitrate, phosphate, light and carbon are
 resources regulating phytoplankton growth \cite{Hu61, Roy}. The reasons of this fact are poorly understood 
although the problem rests under a great attention of ecologists (see, for example, \cite{Moll, Harb,  Record}, among many others) and 
they have suggested numerous different approaches to that problem based on game theory, chaos, stochastics, space inhomogeneities, turbulence etc.

In this paper,  we consider a model of
 an ecological system where  many species share
the very same resource. Our dynamical equations are close to eqs. considered in
\cite{HuWe99}  but we extend the model of \cite{HuWe99}  to take into account
species extinctions and self-limitation effects  (which are important, in particular, for plankton populations, \cite{Roy}).
There is a number of works devoted
to evolution in a multi species context as food-webs, for example,  \cite{Loreau, Loe, Drossel, McKane} as well as the effect of species extinctions  \cite{Kondo, Gall}. These papers have used the Lotka Volterra model where species
interactions slowly evolve in time (that may be connected with foraging \cite{Kondo} or an adaptation of species behaviour) and species extinctions are possible when the abundances attain a critical threshold.
In our model,  all parameters are random but, opposite to \cite{Kondo, Gall},
fixed in time and the primary effect on evolution is species extinctions.

The main results are as follows. We introduce a specific numerical characteristics, which we will call the stress parameter $P_{stress}$.   That characteristics is a natural dimensionless
multiplicative combination of some main parameters involved in the  model formulation, namely, 
the resource turnover rate, the maximum supply of the resource,  self-limitation coefficient and averaged specific growth rates. 
The stress parameter appears, in a natural way, as a result of a  model rescaling and it can be interpreted as a magnitude of selection pressure on ecosystem species
induced by the interaction between the ecosystem and its environment. 
   For example, $P_{stress}$ is large if the resource amount or the resource turnover rate is small.  For large values of the stress parameter,
the model exhibits an effect of convergence to similarity as in the
 competitive Lotka-Volterra systems studied earlier in \cite{Sim}. In contrast to
\cite{Sim}, we obtain a complete  analytical description of the system behaviour.  We  show that the model with or without extinction thresholds
are sharply different. Namely, the simpler model without critical extinction threshold exhibits a global stability, when all positive trajectories converge to the same equilibrium independent of initial state.   The model with extinctions dynamics is fundamentally non-predictable. The trajectories tend to different equilibria, and these final states depend on initial data.  The  biodiversity level,
which we observe after a long evolution, can be expressed via the initial number of species, the stress parameter and other ecosystem characteristics in an explicit way.
This convergence to similarity in a species trait  can be called a "concentration effect". However, this concentration effect does not mean that all
species are completely identical: other traits defined by species parameters can still differ.

The  concentration effect leads to interesting phenomena.  Let us suppose
that initially an ecosystem contains a number of species with random parameters and  consider the limit system, which is a result of a long evolution.   We compare
how a sharp change of environment can affect initial and final systems. By our analytical relations one can estimate the number of species that may go to extinct. The limit system is  more stable than the initial one.

The limit ecosystem, arising as a result of a long evolution, has interesting properties:  its dynamics is governed  by a simple differential equation of the second order. This equation  describes a nonlinear oscillator with a friction and a memory.   If the friction is small and the memory is negligible the dynamics of this oscillator is defined by a Hamiltonian system.
Formation of this universal limit system is a result of species extinctions and a selection pressure on some species parameters, namely those important for species
survival.

\section{Population dynamics}  \label{model}

We consider the following system of equations:
\begin{equation}
     \frac{dx_i}{dt}=x_i (- r_i  + \phi_i(v) -  \gamma_{i} \; x_i), \quad i=1,\dots, M
    \label{HX1}
     \end{equation}
\begin{equation}
     \frac{dv}{dt}=D(S_0 -v)   -  \sum_{i=1}^M c_i \; x_i \; \phi_i(v),
    \label{HV1}
     \end{equation}
where
\begin{equation}
      \phi_i(v)= a_i \Phi(v, K_i), \quad  \Phi(v, K)=\frac{v}{K_i+v}.
\label{MM5}
     \end{equation}
Here $x_i$ are species abundances,    $M$ is the  number of species,  $v$ is  resource amount,
 $D$ is the resource turnover rate, $S_0$ is the maximum supply of resource $v$. In total there are five species specific parameters:  $r_i$ are species mortality,
$c_i >0$ is the fraction of resource consumption by individuals of the $i$ species, $\gamma_{i} >0$ defines species specific
self-limitation, the coefficients $a_i >0$ are species
specific growth rates, and $K_i>0$ are species specific resource constants indicating reduction of resource effect by half.
For $\gamma_{i}=0$  this system have been used to study the  plankton paradox \cite{HuWe99}.  Following \cite{Roy} we assume
$\gamma_i >0$ since it is known that self-limitation is essential for large ecosystems \cite{Alles1, AllPR} and
that plankton and plant ecosystems can induce effects leading to self-limitation \cite{Roy}.
We complement the system (\ref{HX1}),  (\ref{HV1}) with the initial conditions
\begin{equation} \label{Idata}
x_i(0)=\bar x_i, \quad   v(0)=v_0.
\end{equation}

\section{Global stability for  model without extinctions}  \label{Asym}

Here we show that the Cauchy problem
(\ref{HX1}), (\ref{HV1}) and (\ref{Idata}) has a positive solution for all positive Cauchy data. Furthermore, we study both the stability and large time behavior of solutions.

{\bf Proposition I}. {\em
Solution $(x(t), v(t))$ of (\ref{HX1}), (\ref{HV1}) with  initial data
$v(0)\ge 0, \bar x_i=x_i(0) \ge 0$ is defined for all positive $t$,  and it satisfies the estimates
\begin{equation}
 0\le x_i(t) \le \frac  {  \bar x_i \exp(\bar a_i t) }{1 +  \bar x_i \gamma_i \bar a_i^{-1}(\exp(\bar a_i t)  -1)},
\label{estx}
     \end{equation}
where $\bar a_i =a_i - r_i$, and
\begin{equation}
 0\le v(t) \le    S_0 (1- \exp(-Dt)) +  v(0)  \exp(-Dt).
\label{estv}
     \end{equation}
}

{\bf Proof}.  Since  $\phi_i(v) < a_i$, we have $x_i(t) \le y_i(t)$, where
$y_i(t)$ is the solution of the Cauchy problem
$$
\frac{dy_i}{dt}=y_i(-r_i + a_i  - \gamma_i y_i),     \quad y_i(0)=\bar x_i.
$$
Solving this equation, we obtain (\ref{estx}). Estimate  (\ref{estv}) follows from the non-negativity of the term $\sum_{i=1}^M c_i \; x_i \; \phi_i(v)$.

\subsection{Global stability }\label{equ}

Let
\begin{equation} \label{sur66a}
X_i(v)=(\Phi(v, K_i) -p_i)_{+},\;\;\;p_i=r_i/a_i,
\end{equation}
where
$z_{+}=\max\{z, 0\}$, and let also
$$
F_M(v)=F_{M}(v, b, K, p)=\sum_{i=1}^{M} R_i (v, b, K, p),
$$
where
\begin{equation}
R_i(v, b, K, p)=b_i    \Phi(v, K_i)  X_i(v),\;\;\;b_i= c_i \gamma_i^{-1}  a_i^{2}.
\label{RRR}
     \end{equation}
Here the quantity $R_i$  can be interpreted as a consuming rate of $i$th species
and  $F_M$ is the sum of all consuming rates.

Here $a=(a_1,..., a_{M})$, $b=(b_1,..., b_{M})$, $p=(p_1,..., p_{M})$ and $K=(K_1,..., K_{M})$.
If $v$ is a non-negative root of the equation
\begin{equation}
  D(S_0 -v)=F_{M}(v, b, K, p),
\label{Starvd}
     \end{equation}
then
\begin{equation} \label{sur66}
x_i=a_i \gamma_i^{-1} X_i(v), \;\;\;i=1,\ldots,M
\end{equation}
 and $v$ is an equilibrium point of the system
 (\ref{HX1}), (\ref{HV1}).  We  assume here and in what follows that
\begin{equation}\label{K1}
\max_i\,(\Phi(S_0,K_i)-p_i)>0.
\end{equation}
Then the function $F_M$ is non-negative for $v\geq 0$, $F_M(v)=0$ and $F_M(S_0)>0$ due to (\ref{K1}). This implies that equation (\ref{Starvd}) has a unique non-negative solution, which belongs to the interval $(0,S_0)$. We denote this solution by $v_{eq}$.

Let us rewrite equation (\ref{Starvd}) in the following way. Consider first the relation
$$
v=S_0-\frac{1}{D}\sum_{i=1}^{M}b_i\Phi(v,K_i)X_i(w)=G(v, w),
$$
with $w\in [0,S_0]$. Since $G(v,w)$ is decreasing in $w$ from $S_0$ to something which is smaller than $S_0$, for each $w$
the above equation has a unique solution $v=V(w)$. One can verify that the function $V$ is non-decreasing and continuous, $V(0)=S_0$  and $V(S_0) >0$.  We can consider (\ref{Starvd}) as the following fixed-point equation
\begin{equation}
  v=V(v),\;\;v\in [0,S_0].
\label{Starvd7}
     \end{equation}

     In order to describe large time behavior of the system (\ref{HX1}), (\ref{HV1}), we consider the following iterative procedure of solving (\ref{Starvd7}):
 $$
 v^{(k+1)}=V(v^{(k)}), \;\;k=0,1,\ldots,\;\;\mbox{and}\;\;v^{(0)}=0.
 $$
 Then $v^{(1)}=S_0$ and $v_2$ is the solution to
 $$
 v^{(2)}=G(v^{(2)}, S_0),
 $$
 which is positive due to assumption (\ref{K1}). Since $V$ is non-decreasing, we have
 $$
 0=v^{(0)}<v^{(2)}\leq v^{(4)}\leq\cdots\;\;\;\cdots \leq  v^{(3)}\leq v^{(1)}= S_0.
 $$
 We put
 $$
 \hat{v}=\lim_{k\to\infty} v^{(2k)}\;\;\;\mbox{and}\;\,\;\check{v}=\lim_{k\to\infty} v^{(2k+1)}.
 $$
 Clearly,
 \begin{equation}\label{K28a}
 0 < v^{(2)} \leq \hat{v}\leq v_{eq}\leq \check{v}\leq S_0.
 \end{equation}
 Moreover,
 \begin{equation}\label{K28b}
 \hat{v}=S_0-\frac{1}{D}\sum_{i=1}^{M}b_i\Phi(\hat{v},K_i)X_i(\check{v})
 \end{equation}
 and the same relation holds if $\hat{v}$ and $\check{v}$ are exchanged.  Now we can formulate our main result about the large time behavior of solutions to (\ref{HX1}), (\ref{HV1}).


  \vspace{0.2cm}
 {\bf Theorem I}
{\em Let $(x(t), v(t))$ be a solution of (\ref{HX1}), (\ref{HV1}) with positive initial data. Then
\begin{equation}\label{K27a}
\liminf_{t\to\infty}v(t)\geq \check{v},\;\;\limsup_{t\to\infty}v(t)\leq \hat{v}
\end{equation}
and
\begin{equation}\label{K27b}
\liminf_{t\to\infty}x_i(t)\geq X_i(\check{v}),\;\;\limsup_{t\to\infty}x_i(t)\leq X_i(\hat{v}),
\end{equation}
$i=1,\ldots,M$.}

\vspace{0.2cm}
For the proof of this theorem see Appendix.

\vspace{0.2cm}

Note that $\hat{v}=\check{v}$ if $dV/dw>-1$, which is true when, for example, $D$ or $\gamma_0=\min_{i} \gamma_i$, are sufficiently large. Indeed, if
$dV/dw \in (-1, 0]$, the operator $v \to V(v)$ defined on $[0, S_0]$ is a contraction and therefore the  iterations $v^{(k)}$ converge to the same limit. This observation
implies the following

{\bf Corollary I}

{\em For sufficiently large $D>0$ or $\gamma_0>0$
all the solutions $(x(t), v(t))$ of (\ref{HX1}), (\ref{HV1}) with positive initial data
converge, as $t \to \infty$, to   the unique  equilibrium point
defined by eqs.  \eqref{Starvd} and \eqref{sur66}. }


\subsection{Local stability}

Consider now the  problem of stability of equilibrium  states $(x_1,\ldots,x_M,v_{eq})$ defined by  (\ref{Starvd}), (\ref{sur66}).
Denote by ${\bf I}_{eq}$ the set of indices $i$ for which $\phi_i (v_{eq}) -r_i >0$ and by $N_{eq}$ the number of such indices. Then $x_i>0$ when $i\in{\bf I}_{eq}$.

One can show that the eigenvalues of the linear approximation
    of (\ref{HX1}),   (\ref{HV1}) at the equilibrium point $(x_1 ,\ldots,x_M,v_{eq})$ satisfies the equation (see Appendix):
    \begin{equation} \label{eigen}
    \lambda+D+ G(\lambda)=0,
    \end{equation}
where
$$
G(\lambda)=\sum_{i\in{\bf I}_{eq}} c_i\big(x_i\phi_i'(v_{eq})+  \phi_i(v_{eq})\frac{x_i\phi_i'(v_{eq})}{\lambda+P_i(v_{eq})}\big)
$$
and $P_i(v)=\phi_i(v) -r_i$.

Let us show that  $Re \lambda < 0$. In fact, taking the complex conjugate to \eqref{eigen} and
summing these equations we  have
\begin{equation} \label{eigen1}
Re\lambda+D+ Re \ G=0,
 \end{equation}
where
$$
Re \ G=\sum_{i\in{\bf I}_{eq}}c_i\big(x_i\phi_i'(v_{eq})+ \phi_i(v_{eq})\frac{x_i\phi_i'(v_{eq})(Re\lambda+P_i(v_{eq}))}{|\lambda+P_i(v_{eq})|^2}\big).
$$
 This implies that
$$
Re \ \lambda\leq - D-\sum_{i\in{\bf I}_{eq}}c_ix_i\phi_i'(v_{eq})\;\;\mbox{or}\;\; Re \ \lambda\leq -\min_{i\in{\bf I}_{eq}} P_i(v_{eq}).
$$
Thus the equilibrium point $(x_1,\ldots,x_M,v_{eq})$ is locally stable for  all $D$.

\section{Extinctions}  \label{Evol}

System (\ref{HX1}), (\ref{HV1})  does not take into account  species extinctions due to extinction thresholds.  Here we present a model  describing this effect. The system thereby handles the evolution  to the final set of species.
We follow \cite{KVV} with essential simplifications since we do not take into account the emergence of new species.
We start from random values of the model parameters.

Main parameters of our model in this section are  the coefficients $ r_i,  K_i,  a_i$ and  $\gamma_i$. Let us introduce the vector parameter ${\bf P}_i=(r_i,  K_i,  a_i,   \gamma_i)$. Note that $c_i$ is a species specific parameter not necessary to include in this analysis and assumed to be fixed.

Let ${\bf P}=(P_1, P_2, P_3, P_4)$ be a random vector with a probability density
function $\xi ({\bf P})$. This means that the values ${\bf P}_i$ are defined by random sampling, i.e.,
the parameters of the  species are  random independent vectors
${\bf P}_i$  that are drawn from the cone ${\bf R}^4_+=\{{\bf P}: \  P_1 > 0, P_2 > 0, P_3 > 0, P_4 > 0 \}$ by the density $\xi$. Our  assumption to $\xi$ can be formulated as follows:

{\bf Assumption I}.  {\em The probability density function $\xi$ is a continuous function with a support, which has a compact closure   in the positive cone ${\bf R}^4_+$.}

The function $\xi$ is positive on $S_{\xi}$, where $S_\xi$  is an open and bounded set.  The closure   of $S_{\xi}$
we denote by $\bar S_{\xi}$.
Assumption I implies that the mortality rates do not approach  zero and resource consumption is restricted.
It is supposed that initial data $\bar x_i=x_i(0)$ are random mutually independent numbers
		drawn according to a density distribution
		\begin{equation}
     \bar x_{i}  \in  {\mathcal  X}(\bar X,
		 \sigma_{X})
\label{survX1}
\end{equation}
with the mean $\bar X$ and the deviation $\sigma_{X}$.
The random assembly of the species defines an initial state of the ecosystem for $t=0$.

 In order to describe species extinction  we introduce
 a small positive parameter $X_{ext}$ being an extinction threshold. We represent the set of indices $I_M=\{1,2,..., M\}$
as a union of the two disjoint sets:
$$
  I_M=S_e(t) \cup  S_v(t) \quad t\ge 0.
$$
 Here
$ S_e(t)$ is the set of indices of species which exist at the time $t$, and
$ S_v(t)$ is the set of indices of species, which   have disappeared by the moment $t$. Let $N(t)$ denote the number of species in $S_v(t)$ at the moment $t$,
$N(0)=M$.
We assume that  $S_e(0)=\{1,2,..., M\}$,
 and $S_v(0)=\emptyset$.
In our model the species with abundance $x_k$
	vanishes at the moment $t_*$ if
		$x_k(t_*)=X_{ext}$  and
$x_k(t) > X_{ext}$ for  $t < t_*$. The parameter $X_{ext}$ can be interpreted as a threshold for species abundances.

The time evolution of the sets $ S_e(t)$ and $ S_v(t)$
		can be described as follows.

({\bf A})	 if  the $k$-th species
		vanishes at a certain moment $t_*$, i.e. $k \in S_e(t)$ for $t < t_*$ and $x_k(t_*)=X_{ext}$,
			then the index $k$ moves from $S_e(t)$ to $S_v(t)$ at this moment
			$t=t_*$ and we put $x_k(t)=0$ for all $t > t_*$;

({\bf B})	 we assume that the evolution stops at the moment $t_{end}$, if
			at this moment $S_e(t)=\emptyset$.

With modifications described above,  eqs. (\ref{HX1}),  (\ref{HV1})  define the dynamics as follows. Within each time interval $(t_*, T_*)$ between the subsequent species extinctions  the dynamical evolution of $x_i(t)$ is defined by the system
(\ref{HX1}), (\ref{HV1}).

 The quantity $N(t)$ is a piecewise constant decreasing function, therefore,
there exists a limit
\begin{equation} \label{limit}
N(t) \to N_{f},  \quad t \to +\infty
\end{equation}
where $N_{f}$ is the number of species, which survived to the limit state (note that it is possible that $N_{f}=0$).

Let us introduce the parameters
\begin{equation}  \label{delta}
\delta_i= X_{ext}\gamma_i  /a_i
\end{equation}
and assume that
\begin{equation} \label{condeq}
\Phi(S_0, K_i) >  \rho_i=p_i  +  \delta_i,
\end{equation}
for some $i$. Condition (\ref{condeq}) means that the resource supply  is  large enough for existence of a positive equilibrium.

\section{Dynamics of the model with extinctions}  \label{AsymExt}

By \eqref{limit}  there exists a time moment $T_f$  such that
all extinctions have occurred and thus we can use Theorem I and its corollary for the remaining species.  According to section \ref{Evol}
$S_e(T_f)$ is the set of indices corresponding to the  species, which exist for all $t >0$. That set contains $N_f=N(T_f)$ indices.   We modify equation \eqref{Starvd} as follows:
\begin{equation}
  D(S_0 -v_{eq})=F_{ext}(v_{eq}, b, K, p),
\label{StarvdM}
     \end{equation}
where
\begin{equation}  \label{FNeq}
F_{ext}(v, b, K, p)=\sum_{i\in S_e(T_f) } R_i (v, b, K, p).
\end{equation}

 \vspace{0.2cm}
 {\bf Corollary II}
{\em  For sufficiently large $D>0$ or $\gamma_0>0$ all the solutions $(x(t), v(t))$ of (\ref{HX1}), (\ref{HV1}) with positive initial data
converge, as $t \to \infty$, to   an  equilibrium point
defined by eqs.  \eqref{sur66}, \eqref{StarvdM}, and \eqref{FNeq}. That equilibrium depends on the set of remaining species $S_e(T_f)$.}
\vspace{0.2cm}

The assertion follows from the arguments at the beginning of this section and Corollary  I.
 \vspace{0.2cm}

Note that the set $S_e(T_f)$ depends on initial data, therefore, in contrast to Theorem I, we have a number of possible final equilibria.
To show this, let us consider the following situations.  Let $M=3$ and for $X_{ext}$ all three species survive, thus,
$N_{eq}=3$.

Let $X_{ext}>0$ and $x_3(0) = X_{ext} +\kappa$, where $\kappa >0$ is a small number.  We assume that
$x_1(0)-X_{ext}$ and $x_2(0)-X_{ext}$ are not small. Suppose moreover that
$D(S_0 - v(0)) - F_M(v(0)) < 0$ and $|D(S_0 - v(0)) - F_M(v(0))| >> \kappa$. Then it is clear that $x_3$ will go extinct within a short time period
and thus $3$-th species is not involved in the set $S_e(T_f)$ of final equilibria.  If $\kappa$ is not small, then the set contains the $3$-th species.

\section{Concentration of species traits}

Let us consider  the case of arbitrary parameter values, supposing that
the initial number of species  $M >> 1$. For each  $\epsilon >0$ let us denote by $W_{\epsilon}(z)$  the set
of the points  in $\bar S_{\xi}$, which lie in the ball of radius $\epsilon$ centered at  $z=(r,  K, a,   \gamma)$.
The $\epsilon$-neighborhood $W_{\epsilon}(B)$ of a subset $B \subset \bar S_{\xi}$  is the union of $\epsilon$-neighborhoods $W_{\epsilon}(z)$
taken over all the points $z\in B$.

In the set   $\bar S_{\xi}$ we introduce the partial order $\le_e$: $(r_i,K_i,a_i, \gamma_i) \le_{e} (r_j,K_j, a_j, \gamma_j)$ if $a_i \le  a_j$,
$r_i  \ge r_j$, $K_i \ge K_j$ and $\gamma_i \ge \gamma_j$.

Consider the points $z_*=(r_*,K_*, a_*, \gamma_*)$,
which are maximal  with respect to the order $>_{e}$ in the set $\bar S_{\xi}$. Since that set is  closed,
bounded from below
with respect to $K, r, \gamma$, and bounded from upper with respect to $a$, the set $B_*$ of the points
$z_*$ is not empty. It is clear that $B_* $ is a subset of the boundary $\partial S_{\xi}$.
 \vspace{0.2cm}

{\bf Theorem III} (Concentration of traits).
{\em Let Assumption I and (\ref{condeq}) hold and $\epsilon >0$ be a  number.
Then the parameters $a_i,r_i, \gamma_i$ and $K_i$ of species  $x_i$ such that $x_i(t) > X_{ext}$ for all $t>0$  lie in the domain
$W_{\epsilon}(B_*)$ with the probability $Pr_M(\epsilon)$ such that
$Pr_M(\epsilon) \to 1 $ as $M \to +\infty$.
}
\vspace{0.2cm}

{\bf Proof} can be found in Appendix.

If the set $B_*$ is a singleton (i.e. consists of a single point), then we have the concentration trait effect, i.e., all essential parameters of ecosystem
become almost identical as a result of  extinctions. Note that the set $B_*$ is a singleton in the case when
$S_{\xi}$ is a box, i.e.,
$$
S_{\xi}=\{ a_{-} < a_i < a_{+}, \ r_{-} < r_i < r_{+}, K_{-} < K_i < K_{+}, \gamma_{-} < \gamma_i < \gamma_{+}\}.
$$
The set $B_*$ can have a more complicated structure, it may be a union of isolated points or a curve. Also note that even in the singleton case species may differ in coefficients
$c_i$.

\section{Limits of biodiversity in stress environment}  \label{Final}

The following  assertion gives us an information on limits of biodiversity for arbitrary parameter values and our  results  are valid for arbitrary system dynamics: we do not
use here no assumptions on existence of  globally attracting equilibria. Remind that $N_f$ is the number of species, which survive as $t \to +\infty$, i.e.,
the corresponding abundances $x_i(t) \geq X_{ext}$ for all $t \ge 0$.

{\bf Proposition II} {\em The number $N_{f}$  is   bounded by a constant independent of $M$, namely
\begin{equation} \label{Nest}
N_{f} < N_{max}  =\Big[ \frac{DS_0} {X_{ext} a_0 c_0 (p_0+ \delta_0)} \Big] +1,
 \end{equation}
where $[x]$ denotes the integer part of $x$, $c_0=\min c_i$, and
$$
a_0=\min_{r, a, K , \gamma \in \bar S_{\xi}} a, \quad \delta_0=\min_{r, a, K , \gamma \in \bar S_{\xi}} \delta, \quad p_0=\min_{r, a, K , \gamma \in \bar S_{\xi}} p.
$$
}

{\bf Proof}.  First we use an idea from \cite{Hofbauer}.
Let $\langle F \rangle_T = T^{-1} \int_0^T  F(s)ds$ be the average of a  function $F$ on $[0, T]$.   The average of $F$ on $[0, +\infty)$ we  denote by $\langle F \rangle$.
By averaging of \eqref{HV1}   one obtains
\begin{equation}
     T^{-1}\big(v(T)- v(0)\big) =D(S_0 - \langle v\rangle_T )   -  \sum_{i=1}^M c_i a_i \; \langle x_i \Phi(v, K_i) \rangle_T \;.
    \label{HV1A0}
     \end{equation}
  Since the left-hand side here tends to $0$ as $T \to +\infty$ eq. (\ref{HV1A0}) leads to
\begin{equation}
    D(S_0  -\langle v \rangle ) =  \sum_{i=1}^M c_i a_i \; \langle x_i \Phi(v, K_i) \rangle\;
    \label{HV1A1}
     \end{equation}
that in turn entails the estimate
\begin{equation}\label{K27c10}
N_{f} a_0 c_0 X_{ext} \langle \Phi({v},K_*) \rangle < DS_0,
\end{equation}
 where $K_*=\max_{i\in S_e(T_f)} K_i$. Consider the equation in (\ref{HX1}) with the index $i\in S_e(T_f)$ for which $K_i=K_*$. Dividing both sides there by $x_i$, averaging and using that $x_i$ is bounded and separated from zero by $X_{ext}$, we get
 $$
 \langle\phi_i(v)\rangle-r_i=\gamma_i\langle x_i\rangle \geq \gamma_i X_{ext}.
 $$
 Hence
\begin{equation}\label{K27c11}
\langle \Phi({v},K_*) \rangle \geq  p_0 + \delta_0.
\end{equation}
This together with (\ref{K27c10}) leads to (\ref{Nest}).

\bigskip
To find more precise estimates we assume that coefficients $c_i, a_i, \gamma_i$ and $r_i$ satisfies
\begin{equation} \label{assum100}
C_- a<  a_i  <C_{+} a, \quad C_- c <  c_i  <C_{+} c,  \quad 1 \le i \le M,
\end{equation}
\begin{equation} \label{assum101}
C_- \gamma < \gamma_i  <C_{+} \gamma, \quad C_- r <  r_i  <C_{+} r, \quad 1 \le i \le M
\end{equation}
where $a,  c, \gamma, r$ are characteristic values of the corresponding coefficients, $C_{\pm}$ are positive constants independent of
$M,a,c, \gamma, r$.
Let us introduce the stress parameter by

\begin{equation} \label{Stress}
P_{stress}= \frac{c a^2}{\gamma D S_0}.
\end{equation}

To simplify the statement, we also suppose that $K_i=K$. The general assertion on the trait concentration can be formulated as follows:

{\bf Proposition  III} \label{T2}
{\em Suppose Assumption I  and condition (\ref{condeq})   hold.
Let   $i, j$ be two indices such that the corresponding species abundances $x_i(t), x_j(t)$ satisfy $x_i(t) > X_{ext}, x_j(t) > X_{ext}$ for all $t \geq0$.
Then
\begin{equation} \label{Diff}
| p_j  -  p_i |< C_0 P_{stress}^{-1}  (p_0 +\delta_0)^{-1},
\end{equation}
where $C_0 >0$ does not depend on $a, c, \gamma, r$ and $X_{ext}$.

}

{\bf Proof}.
Consider the species such that $x_i(t)> X_{ext}$ for all $t\ge 0$. The corresponding set of indices we denote by $S_e$.  Averaging equations \eqref{HX1} for the species $x_i$ with
$i \in S_e$ we obtain the following relation:
\begin{equation} \label{aver1}
\gamma_i \langle x_i^2 \rangle = a_i  \langle x_i (\Phi(v, K)  -p_i) \rangle.
\end{equation}
Furthermore, we divide \eqref{HX1} on $x_i$ and average the obtained equation that
gives
\begin{equation} \label{aver2}
\langle x_i \rangle =a_i \gamma_i^{-1} \langle \Phi(v, K)  -p_i \rangle.
\end{equation}
The Cauchy inequality implies $\langle x_i^2 \rangle  \ge \langle x_i \rangle ^2$.
Therefore, \eqref{aver1} and  \eqref{aver2} entail
\begin{equation} \label{aver2a}
 \langle x_i(\Phi- p_i) \rangle \ge a_i  \gamma_i^{-1}
\langle  \Phi - p_i \rangle ^2,
\end{equation}
where, for brevity, we use notation  $\Phi=\Phi(v, K)$. By \eqref{HV1} we have
\begin{equation} \label{aver3}
DS_0 \ge  \sum_{i \in S_e} c_i a_i \langle x_i \Phi \rangle .
\end{equation}
We observe that
$$
\langle x_i \Phi \rangle= \langle x_i \Phi - x_i p_i + x_i p_i\rangle.
$$
By the above identity and \eqref{aver1} one has
$$
\langle x_i \Phi \rangle= \langle x_i (\Phi -  p_i)  \rangle + a_i \gamma_i^{-1} p_i \langle\Phi -  p_i \rangle.
$$
That relation and \eqref{aver2a}, \eqref{aver3} lead to the inequality
 \begin{equation} \label{aver4}
DS_0 \ge \langle  \Phi  \rangle \sum_{i \in S_e} c_i a_i^2 \gamma_i^{-1} \langle  \Phi -p_i \rangle 
\end{equation}
that, by \eqref{K27c11}, can be rewritten as follows:
 \begin{equation} \label{Ueq}
P_{stress}^{-1} \ge \langle  \Phi  \rangle \sum_{i \in S_e} \beta_i  \langle  \Phi -p_i \rangle,
\end{equation}
where $\beta_i= c_i  a_i \gamma_i(ca\gamma)^{-1}$ are bounded coefficients independent of $a, \gamma, c$.
Estimate  \eqref{Ueq} entails
\begin{equation} \label{Ueq2}
  (\Phi(v,  K)  -  p_l)_{+}  < C_2 P_{stress}^{-1}\langle  \Phi  \rangle^{-1} \quad \forall \  l\in S_{e}
\end{equation}
 for some $C_2>0$, which is independent of $\gamma, a, c, r$. By (\ref{aver2}) $\Phi-p_i$ is positive and hence  the index ${+}$ in (\ref{Ueq2}) can be removed.
  Combining  \eqref{Ueq2} for $l=i$ and $l=j$ and taking into account that $\langle  \Phi  \rangle > p_0 +\delta_0$ one has \eqref{Diff}.
\vspace{0.2cm}

    Let us derive an estimate of $N_f$ via  $P_{stress} >> 1$ and
the average $\langle \Phi \rangle$.
 We suppose that $p_i=p_0 + (i-1) \Delta p$, $\Delta p << 1$ and
$\beta_i =\beta=O(1)$.
Then estimate \eqref{Ueq} implies
 \begin{equation} \label{FinN}
N_{f} \le   2\beta^{-1} P_{stress}^{-1} ((\langle  \Phi \rangle -p_0) \langle  \Phi  \rangle)^{-1}.
\end{equation}

\begin{figure}[h!]

\includegraphics[width=80mm]{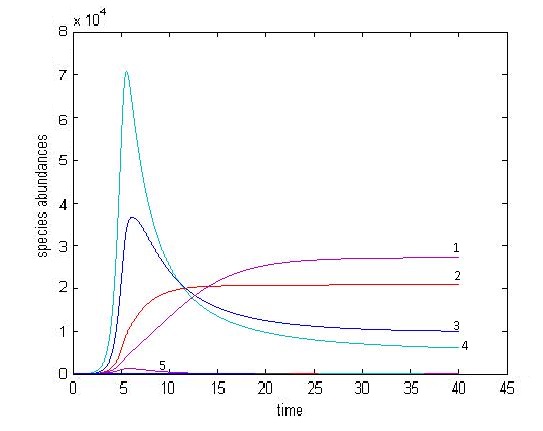}
\caption{\small Dynamics of large  population with a very small $\gamma$. The graphs of  the species abundances $x_i(t)$,   the species number $M=50$. Parameters are
as follows:
$K=4$, $D=10$, $S=100$, $E_y=1, s_y=0.3$, $E_a=2, s_a=0.2$ and $E_r=1- \ln(20), s_r=0.1$.
and $\gamma=10^{-5}$. Here $4$ species coexist  instead  of a single one (they are indicated by numbers $1-4$).}
 \label{Fig2}
\end{figure}

This  calculation is  consistent with numerical simulations. For $\gamma=0.00001$, when all other parameters
have the order $1$,   we obtain a strong concentration effect (see Fig. \ref{Fig2}).  Computations were made for a population of $M=50$ species, where random parameters chosen as explained above.

As a measure of the trait concentration, we can use the quantity
\begin{equation} \label{varp}
Var(p)= max_{i} \{p_i \} - min_{i} \{p_i \}.
\end{equation}
Then the initial $Var(p) \approx 0.6$   but for  $4$  remaining species with large abundances we have $Var(p)=0.07$.
We see that these four species are abundant whereas all other species are extremely rare. Note that these asymptotic results can be generalized in the
case of different $K_i$.

\subsection{Mass extinctions: an analytical approach} \label{Mass}

Relation (\ref{Nest}) allows us to describe, in an analytical way, mass extinctions. Mass extinctions may result
as a consequence of a sharp change of some environmental parameter. It is natural to assume that climate variations or
other abiotic ones can reduce the resource supply level $S_0$. Assume for example that
this reduction is $\Delta S_0>0$ and the new resource supply $S_{new}=S_0 - \Delta S_0$
satisfy 
\begin{equation} \label{rhoV}
S_{new} < X_{ext} a_0 c_0 (p_0 + \delta_0).
\end{equation}
 The last equation implies that for sufficiently large $\Delta S_0$
even all species may go extinct.  We therefore refer this level of the resource $S_{new}$ as a catastrophic level.

Note that  these analytical results show that there are  interesting phenomena. Firstly, let us compare
two ecosystems. One is a random assembly of many species where the variation $Var(p)$ defined by
(\ref{varp})  is large, and  the other ecosystem is a result of long evolution leading to the concentration, i.e.,
$Var(p)$ is small. We find that the concentrated system is more stable with respect to
variations in the resource. Namely,  a sharp change of $S_0$ will kill many more species in the first ecosystem
then in the second one. Secondly, assume  that catastrophes do occur several times yet with a fairly long time in between. Each catastrophe will reduce
the biodiversity yet with less and less probability since the concentration effect becomes stronger and stronger.

To investigate more realistic situations when $K_i$ are different and the parameters are random,   we performed numerical
simulations described in the following section.

\section{Numerical simulations}  \label{Num}

In numerical simulations,  the parameters are chosen as follows.  The coefficients $a_i$ are independent and identically distributed (i.i.d.)
random quantities  such that each $\ln a_i$ is a normally distributed number with the mean $E_a=1$ and the standard deviation $ \sigma_a=0.2$.
This means that each $a_i$ has the same log-normal distribution, $a_i \in \ln {\bf N} (E_a, s_a)$.

Similarly,  the coefficients $r_i$ are i.i.d. random quantities,
 $\ln r_i$ is normally distributed on $[0,1]$ number with the mean $E_r=0.1$ and the standard deviation $ \sigma_r=0.03-\ln(20)$.
The parameters $D=0.1, K=4, S=30$ and $\gamma_i=\gamma=0.001$. The coefficients $c_i$ are random numbers
uniformly distributed on $[0,1]$ and normalized in such a way that $\sum c_i=1$. The initial data $X_i$ are i.i.d. random numbers
distributed log-normally, $X_i =\exp(y_i)$, $y_i \in {\bf N}(E_y, s_y)$  with parameters $E_y=1, s_y=0.3$.
We suppose that all $c_k, r_j$ and $a_l$ are mutually independent.

\begin{figure}[h!]
\includegraphics[width=80mm]{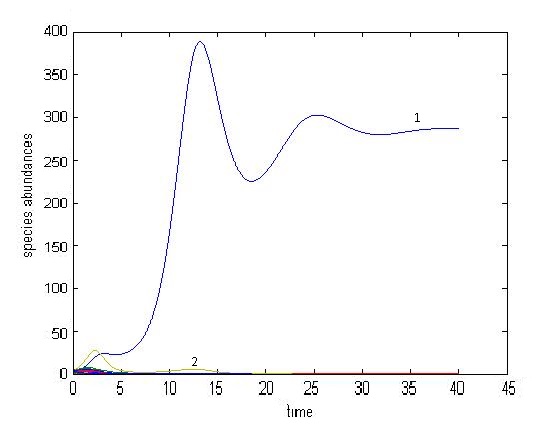}

\caption{\small Dynamics of a species community. The  species number $M=50$, $D=0.1$ and $S_0=30$.  We observe oscillations and finally  that the competition exclusion  principle works: only a single set of parameters remains. This can imply a single species (indicated by $1$), especially if no other traits are important as assumed in the model. }
\label{Fig1}
\end{figure}

For these random species communities and $N=50$ we observe oscillations and then a convergence to an equilibrium (see Fig. \ref{Fig1}).

\begin{figure}[h!]
\includegraphics[width=80mm]{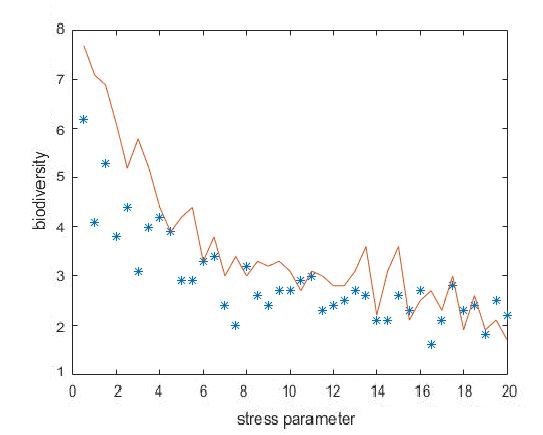}
\caption{\small  The  species number $M=100$.  The species parameters $K$ and $p$ are random numbers obtained by log-normal distributions,
$K=\exp( \tilde K), p=\exp(\tilde p)$, where $\tilde K \in {\mathcal N}(K_0, \sigma_K)$ and $\tilde Kp\in {\mathcal N}(p_0, \sigma_p)$.
The star curve corresponds to the case  $K_0=1, \sigma_K=0$,  $p_0=-1, \sigma_p=0.2$.  For the continuous curve the parameters are the same but
we have a variation in $K$:  $\sigma_K=0.5$.
}
\label{FigBio}
\end{figure}

By simulations we have considered the dependence of biodiversity and concentration trait effect on the stress parameter for populations with random
parameters $K_i, p_i$.  The results are consistent with analytic considerations of the previous section and can be illustrated by  Fig. \ref{FigBio}.

In the numerical simulations of biodiversity we can also observe the trait concentration. For the example illustrated by Fig. \ref{FigBio} the variation  $Var(p)$ 
decreases very strongly as a result species extinctions and this reduction increases as the stress parameter increases.

\section{Dynamics of limit ecosystem}  \label{Lim}


According to Theorem III, if the set $B^*$ consists of a single point, for $M >>1$
the limit system (that appear as a result of many extinctions) has the property $p_i \approx p, K_i \approx K$, where
$p, K$ are some parameter values. To understand dynamics of that system, we consider system (\ref{HX1}), (\ref{HV1}) in the case
$p_i=p, K_i = K$.
 Let us introduce  a new variable
$ Q=-p t + \int_0^t  \Phi(v(s), K) ds.
$
The variable $Q$ is an analogue of "quality of life"  introduced in \cite{Volterra} for the linear case
$\Phi(v, K)=v$.  This case is  studied in \cite{Koz1}. Results of \cite{Koz1} can be extended
to our limit model.  We seek solutions to eqs. (\ref{HX1}) in the form
$$
x_i(t)=C_i(t)  \exp( a_i Q(t)),
$$
where $C_i$ are new unknowns. From (\ref{HX1})  one obtains
$$
\frac{dC_i}{dt}=-\gamma_i C_i^2   \exp( a_i Q(t)),  \quad C_i(0)=\bar x_i,
$$
that gives
$$
\frac{dC_i}{dt}=-\gamma_i C_i^2   \exp( a_i Q(t)).
$$
By solving these equations, we find
$$
C_i=\frac{ C_i(0)   }{1+ \gamma_i C_i(0) \int_0^t \exp( a_j Q(t')) dt'}.
$$
Using the last relation, by (\ref{HV1}) one obtains
\begin{equation} \label{phi}
v=\frac{K P}{1  - P}, \quad \frac{dv}{dt}=\frac{K }{(1-  P)^2} \frac{dP}{dt},
 \end{equation}
where $P=dQ/dt+p$.

After some straight forward computations eqs.  (\ref{HX1}), (\ref{HV1}) reduce to the system
\begin{equation}
    \frac{K }{(1- P)^2} \frac{dP}{dt}=D (S_0- \frac{K P}{1  - P})   -  P f( Q(\cdot))
  \label{HV1Da}
     \end{equation}
\begin{equation}
    \frac{dQ}{dt}=P-p,
  \label{HV1Db}
     \end{equation}
where
\begin{equation} \label {fQ}
 f(Q(\cdot))=\sum_{j=1}^N \frac{c_j \;  a_j  \bar x_j \exp( a_j Q(t)) } {1+ \gamma_j \bar x_j \int_0^t \exp( a_j Q(t')) dt'}.
\end{equation}

Eq. (\ref{HV1Da})  describes a nonlinear oscillator with a  damping term and  nonlinearities with a time delay. Note that $f$ depends on initial data $\bar x_i$.
So, we see that   the  limit ecosystem  can be considered as   a nonlinear oscillator
with a friction and a memory.  The oscillator state is determined by two variables: $P$ and $Q$. The first variable  is a difference between the normalized  species consuming rate $\Phi(v, K)$ and the normalized species mortality rate, i.e., admits a biological interpretation.
This variable can be called Malthusian parameter.
The second variable $Q$ does not admit a simple explicit
interpretation. It is a generalization of the quality of life introduced by Volterra's.  Note that $Q$ is the integral of $P$, i.e., it can be considered as an integral Malthusian parameter. Since this is a parameter expressing a trait over long time we can call  $P$ the sustainable Malthusian parameter.

Eq. (\ref{HV1Da})  can be simplified  in   two cases: for $\gamma=0$ and for bounded times,  $t <<\ ln(\gamma^{-1})$ (an initial stage) and for $t >> 1/\gamma$ (large times, the final stage).  In the first case from \eqref{fQ} we have
\begin{equation} \label {fQ2}
 f(Q(\cdot))=f(Q(t))=\sum_{j=1}^N  c_j \;  a_j  \bar x_j \exp( a_j Q(t)).
\end{equation}

 We obtain an oscillator, which  is a perturbed  Hamiltonian integrable system without memory.
In this case for small $D$ the solutions of (\ref{HV1Da}) tend to the equilibrium in an oscillating manner  ( see  Fig.
\ref{Fig1}).

\section{Conclusion}

In this paper, we have investigated a model of ecosystems exploiting a single resource and interacting with
the environment.

Until May's seminal works \cite{May1, May2}, ecologists believed
 that large complex ecosystems,  involving
 a larger number of species and interconnections,
are stable.  May \cite{May1, May2} considered
 a community of $S$ species with connectance
$C$ that measures the number of realized links with respect to  the
number of all possible links.
R. May's   analysis of  local stability of an equilibrium
gave  quite revolutionary results that inspired a great discussion.
 It was shown that for large systems with random interaction parameters the instability
can occur for large  $C$. More connected communities are
 more unstable. This  approach is developed in
\cite{Alles1, Alles2}, where more complicated networks with
interactions of different types (predator-prey, amensalism,
mutualism, competition) were studied.

All these fundamental results hold under the assumption that, at an equilibrium, ecosystems have a random structure, namely, 
the entries of the matrix, that defines the linearization of system at the equilibrium,  are distributed according to  smooth densities, for example, Gaussian ones.

In this paper, we use a similar assumption but on the initial choice of  species traits.   The initial distribution of  species  traits is defined by  continuous densities with
non-empty supports, i.e., roughly speaking, the species traits are distributed homogeneously   in a   domain.
We show
that  in the evolution process the distribution of species traits  becomes more concentrated  when ecosystem evolves under a stress or as a result of species extinctions.
During the evolution process  the domain of species trait localization shrinks. That small domain of localization means that species become more and more similar (as in \cite{Sim}).  
In contrast to \cite{Sim}, we do not use any specific assumptions on the adaptation of system parameters. 
We have found a parameter, which
defines the stress level. This parameter depends on the supply level, turnover rate, and resource consuming intensity.

 The most interesting effect  of species trait concentration  is as follows.   For large times
a  stable and simple limit ecosystem appears just because of most species goes
extinct  under stress.
For large times ecosystem dynamics and extinctions  of species under stress
produces a self-organized community consisting of species with close consumer efficiencies (note that these species can be different in other traits). In some cases, dynamics of this limit community can be described by a simple equation, which describes a nonlinear oscillator with a  friction and a memory, which is close to a Hamiltonian system. We have found an asymptotic approach to study this system. Note that the reduction mechanism differs from  previously found one in \cite{Barab, Koz1}. In \cite{Barab}
a mean field approach is applied to complex ecosystems and gene networks. This approach exploits the system topology, when species
(genes) can interact with many others. Complicated systems of equations were reduced to a single differential equation of the first order. Such equations do not exhibit time oscillations
whereas our equation  simulates a perturbed nonlinear oscillator and  it can describe slowly decreasing oscillations.   In  \cite{Koz1}  a reduction to Hamiltonian systems is also based on topological properties
of interactions in ecosystems. So, the  reduced descriptions of  complex systems proposed in \cite{Barab, Koz1} can be called a topological one. In contrast to \cite{Barab, Koz1}, 
in the present paper the reduced description is based not only on the system topology (i.e., the fact that species share the same resource) but also on others phenomena: extinctions and selection by a tough environment (which can be measured by the stress parameter).

These results can be useful for understanding why   ecosystems where species feeds
on few resources can have a large biodiversity, and how mass extinctions depend on environment and ecosystem parameters.  The intriguing effect is that we observe
a picture similar to
statistical physics: the state of a ecosystem which arises as a product of a long evolution can be described by two quantities $P, Q$ having a  biological interpretation.
Namely,  $P$ can be called Malthusian parameter, and this quantity determines a balance between mean mortal and  growth rates.  The second quantity $Q$ can be named
sustainable Malthusian parameter,  and it  can be obtained by integrating $P$ over time.

We have computed analytically the number of finally coexisting survived species and how to this number depends
on main ecosystem parameters (the resource supply, the mortality rates, the resource turnover etc).
It is shown that the main quantity that determines final biodiversity is the stress parameter.

\section{ Acknowledgements}

The authors are thankful to Referee for useful remarks.
The second author was supported by Linkoping University and by the Government of the Russian Federation through mega-grant 074-U01,
 and RFBR grant  16-01-00648.

\section{Appendix}

\subsection{Proof of formula (\ref{eigen})}

Assume that all $P_i(v_{eq})>0$. Then the eigenvalue problem for the linear part of the right-hand side of (\ref{HX1}), (\ref{HV1})
at the equilibrium point $(x_1,\ldots,x_M,v_{eq})$ has the form
\begin{equation}\label{KK1}
-\gamma_ix_iX_i+x_i\phi_i'(v_{eq})V=\lambda X_i,\;\;i=1,\ldots,M,
\end{equation}
\begin{equation}\label{KK2}
-\sum_{i=1}^Mc_i\phi_i(v_{eq})X_i-\big(D+\sum_{i=1}^Mc_ix_i\phi_i'(v_{eq}) \big)V=\lambda V,
\end{equation}
where $(X_1,\ldots,X_M,V)$ is an eigenvector corresponding to the eigenvalue $\lambda$. Solving the first system with respect to $X_i$ and
inserting the solution into the second equation we obtain
$$
X_i=\frac{x_i\phi_i'(v_{eq})}{\lambda+\gamma_i x_i}
$$
and
\begin{equation}\label{KK3}
\lambda+D+\sum_{i=1}^M c_i x_i \phi_i'(v_{eq})+\sum_{i=1}^Mc_i\phi_i(v_{eq})\frac{x_i\phi_i'(v_{eq})}{\lambda+\gamma_ix_i}=0.
\end{equation}
Since $\gamma_ix_i=P_i(v_{eq})$ at the equilibrium point we arrived at (\ref{eigen}).

If some of $P_i(v_{eq})$ are non-positive  the corresponding terms in (\ref{KK1})-(\ref{KK3}) are zeros and we again arrive at (\ref{eigen}).

\subsection{Proof of Theorem I on global stability of   positive solutions} \label{solasym}

  We apply a special method based on the theory of decreasing operators in Banach spaces
(see \cite{Ban} and references therein)  that allows us to prove this assertion without any additional assumptions.
This approach is applicable here due to special properties of monotonicity of our problem.

Let us  rewrite \eqref{Starvd7}
as follows:
\begin{equation}
  \bar v= V (\bar v),
\label{App1}
     \end{equation}
where the operator $V$ is described in subsection \ref{equ}.
We remind that $V (\bar v)$ is a decreasing function in $\bar v$.

Our next step is to rewrite system \eqref{HX1}, \eqref{HV1} as an integral equation for unknown function
$v(t)$. Let $w(t)$ be a given non-negative, continuous, bounded  function on $[0,\infty)$ having a limit $\bar w$ at infinity. We can resolve eqs. \eqref{HX1} (with $v$ replaced by $w$) following  section \ref{Lim}. As a result, we obtain
\begin{equation}
  x_i(t)={\bf X}_i(w(\cdot))(t),
\label{AppX}
\end{equation}
where
$$
{\bf X}_i(w(\cdot))(t)=\frac{x_i(0)}{ J_i(w(\cdot))(t)},
$$
and
$$
 J_i=\exp(- \int_0^t P_i(w(s)) ds) + \gamma_i x_i(0) \int_0^t \exp(- \int_{t_1}^t P_i(w(s)) ds)  dt_1.
$$

One can verify that for $x_i(0) >0$
$$
{\bf X}_i(w)(t)\to X_i(\bar w)\;\;\,\mbox{as $t\to\infty$,}
$$
where $X_i$ is defined in subsect. \ref{equ}

Next, we can solve eq. \eqref{HV1} with respect to $v$, where $x_i$ is given by (\ref{AppX}) and $v(0)=v_0$. We denote this solution by  ${\bf V}(t)={\bf V}(w(\cdot))(t)$.
We cannot write this solution explicitly but we need in what follows only some of its properties. First, this solution is a decreasing function with respect to $x_i$ and consequently with respect to $w$.
Second,
$$
{\bf V}(w(\cdot))(t)\to \bar v \;\;\,\mbox{as $t\to\infty$,}
$$
where $\bar v=V(\bar w)$. Thus the unique solution to the problem \eqref{HX1}, \eqref{HV1} with the Cauchy data (\ref{Idata}) can be  obtained by solving the following fixed point problem
\begin{equation}\label{Appw}
v(t)={\bf V}(v)(t)\;\;\;
\end{equation}
and then
$$
x_i(t)={\bf X}_i(v(\cdot))(t),\,\;i=1,\ldots,M.
$$
To solve the equation $v={\bf V}(v(\cdot))$ in the class of bounded, continuous, non-negative functions (denoted by  ${\bf B}$), we use the following iterations
$$
v_{n+1}(t)={\bf V}(v_n(\cdot))(t),\;\;n=1,2,\ldots,\;\;v_0(t)=0.
$$
Then
$$
v_0\leq v_2\leq v_4\leq\cdots\;,\;\;\;v_1\geq v_3\geq\cdots\;\;\;\mbox{and}\;\;v_{2j}\leq v_{2k+1}\;\;\mbox{for all $j,k$}
$$
(here $\leq$  denotes the partial order  on  ${\bf B}$: $v \leq u$ if $u(t) \le v(t) \ \forall t \in [0,T]$).

To show the convergence of the odd and even iterations, we observe that we can consider the fixed point equation (\ref{Appw}) on a finite interval $(0,T)$.
Now the operator ${\bf V}:C[0,T]\rightarrow C[0,T]$ is compact and hence the odd and even terms of sequences converge on $[0,T]$ for each $T$. We introduce their limits
$$
\check{V}(t)=\lim_{j\to\infty}v_{2j}(t),\;\;\;\hat{V}(t)=\lim_{k\to\infty}v_{2k+1}(t).
$$
Then ${\bf V}(\check{V})=\hat{V}$ and ${\bf V}(\hat{V})=\check{V}$. Let $\hat{x}_i$ be given by (\ref{AppX}) with $w=\hat{V}$ and $\check{x}_i$ be given by (\ref{AppX}) with $w=\check{V}$. Then the vector function $(\check{x}_1,\ldots,\check{x}_M,\hat{V})$ satisfies the problem
\begin{eqnarray*}
&&\frac{d\check{x}_i}{dt}=\check{x}_i (- r_i  + \phi_i(\check{V}) -  \gamma_{i} \; \check{x}_i), \quad i=1,\dots, M\\
&&\frac{d\hat{V}}{dt}=D(S_0 -\hat{V})   -  \sum_{i=1}^M c_i \; \check{x}_i \; \phi_i(\hat{V}),
\end{eqnarray*}
and the functions $(\hat{x}_1,\ldots,\hat{x}_M,\check{V})$ are solutions of
\begin{eqnarray*}
&&\frac{d\hat{x}_i}{dt}=\hat{x}_i (- r_i  + \phi_i(\hat{V}) -  \gamma_{i} \; \hat{x}_i), \quad i=1,\dots, M\\
&&\frac{d\check{V}}{dt}=D(S_0 -\check{V})   -  \sum_{i=1}^M c_i \; \hat{x}_i \; \phi_i(\check{V}),
\end{eqnarray*}
Moreover,  the last two systems have the same Cauchy data. Taking differences we obtain a homogeneous Cauchy problem for $(\hat{x}_1-\check{x}_1,\ldots,\hat{x}_M-\check{x}_M,\hat{V}-\check{V})$ and by uniqueness for the Cauchy problem we obtain that $\check{V}=\hat{V}$.

Let us turn to the asymptotic behaviour of the fixed-point solutions. Let $\bar v_k=\lim_{t\to\infty}v_k(t)$. Then
$$
\bar v_0=0\;\;\;\mbox{and}\;\;\bar v_{k+1}=V(\bar v_k),\;\;k=0,\ldots.
$$
This proves inequalities (\ref{K27a}) and (\ref{K27b})
 and completes the proof of Theorem I.

\subsection{Proof of Theorem III}

 It proceeds in three steps.

{\em Step 1:  Monotonicity of species abundances}.

Consider a point $\bar z=(a_i, r_i, \gamma_i, K_i)$, which are not contained in  $W_{\epsilon}(B_*)$, and the corresponding species
population $x_i(t)$. Suppose that for all
$t \ge 0$ we have
\begin{equation} \label{Xi}
 x_i(t) > X_{ext}.
\end{equation}
Consider $j$-th species with  parameters $(a_j, r_j, \gamma_j, K_j)$ and the species abundance
 $x_j(t)$. We assume that
\begin{equation} \label{XiC}
 x_j(0) \ge x_i(0), \quad  r_i  \ge r_j, \ a_i \le a_j, \gamma_i \ge \gamma_j,
K_i \ge K_j.
\end{equation}
Then
\begin{equation} \label{Xi2}
 x_j(t) \ge x_i(t)  \quad  \forall t >0.
\end{equation}
Indeed, let us consider equations for $x_i, x_j$:
\begin{equation}
     \frac{dx_i}{dt}=x_i (- r_i  + \phi_i(v) -  \gamma_{i} \; x_i),
    \label{HX1i}
     \end{equation}
\begin{equation}
     \frac{dx_j}{dt}=x_j (- r_j  + \phi_j(v) -  \gamma_{j} \; x_j).
    \label{HX1j}
     \end{equation}
If (\ref{Xi2}) is violated then there is a time moment $t_1 >0$ such that
\begin{equation} \label{Xi21}
 x_j(t_1) = x_i(t_1),  \quad  \frac{dx_i}{dt}(t_1) > \frac{dx_j}{dt}(t_1).
\end{equation}
But $$x_i(t_1) (- r_i  + \phi_i(v) -  \gamma_{i} \; x_i(t_1)) \le x_i(t_1) (- r_i  + \phi_i(v) -  \gamma_{i} \; x_i(t_1))$$ due to
the first inequality in (\ref{Xi21}) and  (\ref{XiC}). The last inequality  contradicts the second inequality in
in (\ref{Xi21}), thus,  (\ref{Xi2}) is proved.

Inequality (\ref{Xi2}) shows that if the species  $x_i$ survives for all times, then  all the species with parameters satisfying (\ref{XiC})
also survive for all $t>0$.

{\em Step 2: a priori boundness of biodiversity}. Here we use Proposition II.  The number $N_s$ of species, which survive for all times, a priori bounded by the system parameters and does not depend on
$M$ as $M \to \infty$. We refer the corresponding set of species parameters   as ${\mathcal P}_s$. Due to Prop. II,
\begin{equation} \label{NP}
     N_s <  C,
     \end{equation}
where $C>0$ is independent of $M$.

{\em Step 3}.  Let us consider the $\epsilon$-neighborhood $W_{\epsilon}(B_*)$.  Suppose there exists a point $\bar z \notin  W_{\epsilon}(B_*)$. The initial data $x_i(0)$
for the corresponding species we denote by $\bar x_i$. Then, according to
Step 1,  the set
${\mathcal P}_s$ contains all points $z$ from $W_{\epsilon}(B_*)$ such that $z \ge_e \bar z$. We denote the set of such points
by $W_{\epsilon, \bar z}(B_*)$.  Note that
due to the conditions to the set $S_{\xi}$ (see Assumption I), the set   $W_{\epsilon, \bar z}(B_*)$ contains a small open ball. Therefore,
since $\xi$ is positive on the interior of $S_{\xi}$ (see Assumption I), we have
$$
1> J=\int W_{\epsilon, \bar z}(B_*) \xi(z) dz > \delta_{\epsilon, \bar z} >0.
$$
 The number $\delta_{\epsilon, \bar z}$ is independent of $M$. Consider the event $E=A B$  where $A$ is the event that
the species parameters lie in $W_{\epsilon, \bar z}(B_*)$ and $B$ is the event that
initial data $x_i(0) >\bar x_i \ \forall i$. The events $A$ and $B_s$ are independent and $Prob(A) >0$ due to the above estimate for $J$.
According to hypothesis on the random choice of $x_i(0)$ we also have $Prob(B) >0$.  Therefore, $Prob(E)=q >0$.

Consider the event $E_{M, N_s}$ that among $M$ species there are
  not more than $N_s$ species such that the corresponding species parameters lie in $W_{\epsilon, \bar z}(B_*)$ and that
initial data $x_i(0) >\bar x_i \ \forall i$. The probability of $E_{M, N_s}$ can be computed by the Bernoulli relation, and we have
$$
Prob(E_{M, N_s}) < \sum_{k=0}^{N_s}  M^{k} (k!)^{-1} q^{k} (1-q)^{M- k}.
$$
 We see that  $E_{M, N_s} \to 0$ as $M \to \infty$ and the Theorem III  is proved.



\begin{thebibliography}{9999}

\bibitem{Barab}   J. Gao, B. Barzel, and A-L. Barab\'asi, Nature {\bf 530} 307–312 (2016)

\bibitem{Volterra} V. Volterra,  {\it Lecons sur la theorie mathematique de la lutte pour la vie}, (Gauthier -Villard, Paris, 1931).

\bibitem{Hardin}  G. Hardin,   The competitive exclusion principle. Science, \textbf{ 131}, 1291, (1960)

\bibitem{Til77} D. Tilman, Ecology {\bf 58}, 338 (1977)

\bibitem{Zeeman} E. C. Zeeman and M. L. Zeeman, Transactions of the American Mathematical Society, {\bf 355} ,  713 (2003)

\bibitem{Hu61} G.E. Hutchinson, Am. Nat. {\bf 95}, 137 (1961).

\bibitem{Roy} S. Roy, J. Chattopadhyay,  Ecological complexity, {\bf 4} 26 (2007) 

\bibitem{Moll} J. D. Moll and J. S. Brown,  American Naturalist. {\bf 171} 839(2008).

\bibitem{Harb} C.W.  Harbison,   Ecology  {\bf 89} 3186 (2008)

\bibitem{Record} N. R. Record,  A. J. Pershing, , and F. Maps, F.  ICES Journal of Marine Science, {\bf 71} 236 (2014)



\bibitem{HuWe99} J. Huisman and F.J. Weissing, Nature {\bf 402}, 407 (1999).


\bibitem{Loe} N. Loeuille  and  M. Loreau  Proc. Natl. Acad. Sci. USA
{\bf 102}, 5761,  (2005) 

\bibitem{Drossel}  B. Drossel, A. J. McKane,  and Ch. Quince,
 Journ. of Theoretical Biology,  {\bf 229} 539  (2004).

\bibitem{Loreau}  A. Brannstrom, N. Loeuille, M. Loreau, and U. Dieckmann,  {\bf 4} Theor. Ecol. (2011)



\bibitem{McKane}  A. J. McKane   Eur. Phys J.,
{\bf B38} 287–295, (2004)

\bibitem{Kondo} M. Kondoh, Science   {\bf 299},  288  (2003)


\bibitem{Gall} G. J. Ackland and I. D. Gallagher, Phys. Rev. Letters,  {\bf 93}   158701-1 - 158701-4. (2004)

\bibitem{Sim}  M. Scheffer and E. H. van Nes, Proc. Nat. Acad. Sci, vol. 103, no. 16 6233- 6235 (2006)



\bibitem{Alles1} S. Allesina and Si Tang,  Nature, {\bf 483}, 205 (2012).

\bibitem{AllPR} S. Allesina (private communication).


\bibitem{KVV}    V. Kozlov, S. Vakulenko and U. Wennergren,   Bulletin of Math. Biology, {\bf 78} 2186 (2016)

\bibitem{Hofbauer}  J. Hofbauer and      K. Sigmund, {\it Evolutionary Games and Population Dynamics}, (   Cambridge University Press, 1998  ) 


\bibitem{Koz1}    V. Kozlov, S. Vakulenko and U. Wennergren, Hamiltonian dynamics for complex food webs, Phys. Rev {\bf E93} 032413 (2016).


\bibitem{May1} R. May,  Nature(London), {\bf 238}, 413 (1972).

\bibitem{May2} R. May,  {\it Stability and complexity in model ecosystems} ( Princeton Univ. Press,
Princeton, 1974).





\bibitem{Alles2} S. Allesina,  Nature, {\bf 487}, 175 (2012).


\bibitem{Ban} G. Herzog and P. C. Kunstmann,  Numerical Functional Analysis and Optimization,
{\bf 34},  530 (2013).
































\end{thebibliography}

\end{document}